\documentclass[runningheads,a4paper]{llncs}

\usepackage{amssymb}
\usepackage{graphicx}
\usepackage{apacite}
\newcommand{\keywords}[1]{\par\addvspace\baselineskip
\noindent\keywordname\enspace\ignorespaces#1}
\usepackage{caption}
\usepackage{subfigure}
\usepackage{graphicx}
\usepackage{amsmath,amssymb,url}
\usepackage{color}
\usepackage{musicography}
\usepackage[ruled,vlined]{algorithm2e}

\usepackage{calc}
\usepackage[absolute]{textpos}
\newlength{\footerxpos}
\newlength{\footerypos}
\newcommand{\copyrightnote}{\copyright\:2020 Copyright held by the authors. This is the author's version of the work. It is posted here for your personal use. Not for redistribution. The definitive Version of Record will be published in \textit{Proceedings of the 2020 Joint Conference on AI Music Creativity} (CSMC-MuMe 2020), ISBN 978-91-519-5560-5.}

\begin{document}

%
\begin{textblock*}{\columnwidth}(\footerxpos,\footerypos)\noindent\footnotesize{\copyrightnote}\end{textblock*}

\mainmatter  

\title{Automatic Analysis and Influence of Hierarchical Structure on Melody, Rhythm and Harmony in Popular Music}

\titlerunning{Automatic Analysis and Influence of Hierarchical Structure in Popular Music}

%
%
\author{Shuqi Dai \and Huan Zhang \and Roger B. Dannenberg}
%

\institute{Computer Science Department, Carnegie Mellon University \\ \email{shuqid@cs.cmu.edu, huanz@andrew.cmu.edu, rbd@cs.cmu.edu}}

%
%

\maketitle

\begin{abstract}
Repetition is a basic indicator of musical structure. This study
introduces new algorithms for identifying musical phrases based on repetition. Phrases combine to form \textit{sections} yielding a two-level hierarchical 
structure. Automatically detected hierarchical 
repetition structures reveal significant interactions
between structure and chord progressions, melody and 
rhythm. Different levels of hierarchy interact differently,
providing evidence that
\textit{structural hierarchy} plays an important role in music beyond
simple notions of repetition or similarity.
Our work suggests new applications for
music generation and music evaluation.
\keywords{Music Structure, Music Understanding, Structure Analysis, Multi-level Hierarchy, Music Segmentation, Pattern Detection, Repetition, Music Similarity}
\end{abstract}

\section{Introduction}
Form and structure are among the most important elements in music and have been widely studied in music theory. Music structure has a hierarchical organization ranging from low-level motives to higher-level phrases and sections. These different levels 
influence the organization of other elements such as harmony, melody and rhythm, but these influences are not well formalized. 
MIR research has developed techniques for detecting music segmentation and repetition structures, but hierarchy is often ignored 
\cite{Dannenberg2009AcousticS, paulus2010state}. 
A fundamental question 
is whether higher levels of hierarchy are essentially just larger groupings or whether different levels play different roles.
If the latter is true, then a better representation and understanding of hierarchy should
be useful for prediction, generation, analysis and other tasks.
Long-term structure in music is also a recent topic in music generation with deep learning,
and attention models such as the Transformer \cite{vaswani2017attention, huang2018music} seem to improve results. While this
suggests some data-driven discovery of structure, results are hard to interpret and,
for example, it is not clear whether hierarchy plays a role.
. 

We began our study by developing a method to identify low-level structure in popular songs. Our approach identifies phrases with repetition of harmony, melody and rhythm.
Next, we discovered a simple 
way to infer higher-level structure from this phrase-level structure. Beyond viewing structure
as mere repetition, we show that chord progressions, melodic structures and rhythmic
patterns are all related to music structure, and there are significantly different 
interactions at different levels of hierarchy. Our main contributions are: 
1) a novel algorithm to extract repetition structure at both phrase and section levels
from a MIDI data set of popular music, 
2) formal evidence that melody, harmony and rhythm are organized to reflect different levels of hierarchy,
3) data-driven models offering new music features and insights for traditional music theory.
We believe that this work is important in highlighting roles that structure can play in music analysis, music similarity, music generation, and other MIR tasks.

Section \ref{sec:related_work} discusses related work, and Section \ref{sec:structure_analysis} presents our phrase-structure analysis method. 
Section \ref{sec:stucture_exploration} 
describes general properties of structures we found, while Section 
\ref{sec:harmonic_analysis} explores the relationships between structures and harmony, melody and rhythm, respectively. 
Finally, Sections \ref{sec:discussion} and \ref{sec:conclusion} present discussion and conclusions.

\section{Related Work}\label{sec:related_work}
Computational analysis of musical form has long been an important task in Music Information Retrieval (MIR). Large-scale structure in music, from classical sonata form to the repeated structure in pop songs, is essential to music analysis as well as composition. Marsden \citeyear{Marsden2010RecognitionOV} implemented Schenkerian analysis and applied it to Mozart variations.
Hamanaka, Hirata and Tojo \citeyear{Hamanaka2014MusicalSA} describe a tool for Generative Theory of Tonal Music (GTTM) analysis that matches closely the analyses of musicologists.
Allegraud et al. \citeyear{Allegraud2019LearningSF} use unsupervised learning to segment Mozart string quartets. 
Go, Ryo, Eita and Kazuyoshi \citeyear{Shibata2019StatisticalMS} perform structural analysis using homogeneity, repetitiveness, novelty, and regularity. Our work builds on the
idea of extracting structure by discovering repetition. 

Identifying hierarchical structure is likely to play a role in music listening. Granroth-Wilding  \citeyear{Wilding2013Combinatory} employs ideas from Natural Language Processing (NLP) 
to obtain a hierarchical structure of chord sequences.
Marsden, Hirata and Tojo \citeyear{Marsden2013TOWARDSCP} state that advances in 
the theory of tree structures in music
will depend on clarity about data structures and explicit algorithms.
Jiang and Müller \citeyear{jiang2013automated} propose a two-step segmentation algorithm for analyzing music recordings in predefined sonata form: a thumb-nailing approach for detecting coarse structure and a rule-based approach for analyzing the finer substructure. 
We present a detailed algorithm for segmenting music into phrases and deriving a higher-level sectional structure starting with a symbolic representation. 

Segmentation of music audio is a common MIR task with a substantial literature. Dannenberg and Goto  \citeyear{Dannenberg2009AcousticS} survey audio segmentation techniques based on repetition, textural similarity, and contrast. Barrington, Chan and Lanckriet \citeyear{Barrington2009DynamicT} perform audio
music segmentation based on timbre and rhythmical properties. However, MIDI has the advantage of greater and more reliable rhythmic information along with the possibility of cleanly separating melody.  Many chord recognition algorithms exist, e.g.
Masada and Bunescu \citeyear{Masada2018ChordRI} use a semi-Markov Conditional Random Field model. References to melody extraction from MIDI can be found in Zheng and Dannenberg \citeyear{Jiang2019MelodyII} who use maximum likelihood and Dynamic Programming. 
Rolland \citeyear{rolland1999} presents an efficient algorithm for spotting matching melodic phrases, which relates to our algorithm for segmentation based on matching sub-segments of music. Lukashevich \citeyear{lukashevich2008towards} proposes a music segmentation evaluation measure considering over- and under-segmentation. Collins, Arzt, Flossmann and Widmer \citeyear{collins2013siarct} develop a geometric approach to discover inexact intra-opus patterns in point-set representations of piano sonatas. Our work 
introduces new methods for the analysis of multi-level hierarchy in MIDI and investigates the interplay of structure with harmony, melody and rhythm.

\section{Phrase-level Structure Analysis}\label{sec:structure_analysis}
We introduce a novel algorithm based on repetition and similarity to extract structure from annotated MIDI files. Given input consisting of a chord and melody sequence for each song together with its time signature (obtained from MIDI pre-processing), the algorithm outputs a repetition structure. In this section, we will introduce the design motivation, structure representation, details of the algorithm and some evaluation results.

\subsection{Motivation and Representation}
We represent the structure of a song with alternating letters and integers that indicate phrase labels and phrase length in measures (all boundaries are bar lines). We indicate \textit{melodic phrases} (where a clear melody is present, mostly a vocal line or a instrument solo) with capital letters and \textit{non-melodic} phrases with lower-case letters. For example, {\tt i4A8B8x4A8B8B8X2c4c4X2B9o2} denotes a structure where {\tt A8} and {\tt B8}, for example, represent different repeated melodic phrases of length 8 measures. The {\tt B9} indicates a near-repetition of the earlier {\tt B8}, but with an additional measure. In addition, {\tt i} indicates an introduction with no melody and {\tt o} is a non-melody ending. {\tt X} and {\tt x} denote extra melodic and non-melodic phrases that have no repetition in the song. (The first and second occurrence of {\tt X2} in the structure do \textit{not} match. We could have labeled them as {\tt D2} and {\tt E2} but {\tt X2} makes these non-matching phrases easier to spot.) Non-melodic
phrases such as {\tt c} often 
refer to a transition or bridge, while
{\tt X} usually indicates non-repeating phrases or just 
inserted measures.

Songs do not have unique structures. Consider a simple song 
with measures \textit{qrstuvwxqrst}. 
Here, matching letters mean repeated measures, based on overall similarity of chords, rhythm onset times, and a melodic distance function.

We assume that shorter 
descriptions are more ``natural'' \cite{simonandsumner}. 
Therefore, we model structural description as a form of data 
compression, e.g. we can represent this song more compactly
as {\tt ABA} where {\tt A}=\textit{qrst} and {\tt B}=\textit{uvwx}. 
This description requires us to represent 3 phrase symbols
({\tt ABA}) plus the descriptions of {\tt A} (\textit{qrst}) 
and {\tt B} (\textit{uvwx}) for a total of 3 phrases and 8 
constituent measures.  The description length here is 
$h \cdot 3 + g \cdot 8$, where $g$ and $h$ are constants 
that favor fewer phrases and more repetition, respectively. 
We manually tuned the settings to $h = 1.0$ and $g = 1.3$ after experimenting with the training data.  In comparison the 
representation {\tt A}=\textit{qrstuvwxqrst} has a description 
length of $h \cdot 1 + g \cdot 12$, which is longer and therefore
not as good. Extending this idea, 
we define \textit{Structure Description Length} (SDL) for a song 
structure $\Omega$ consisting of one or more repetitions of 
phrases from the set $P$ as

\vspace{-8pt}
\begin{equation}
SDL(\Omega) = h \cdot |\Omega| + g \cdot \sum_{\forall p \in P} avglen(p)
\vspace{-5pt}
\end{equation}
\noindent 
where $avglen(p)$ is the average length of instances of phrase $p$. (Recall that matching phrases need not be exactly the same length.) 
Since there are often many possible structure descriptions, $SDL$
allows us to automatically select a preferred one.

\subsection{Data Pre-processing}\label{sec:data}
We use a Chinese Pop song MIDI data set consisting of 909 
manual transcriptions of audio \cite{wang2020pop909}. We use key and chord labels from audio in combination with labels automatically derived 
from MIDI, resolving differences with heuristics to improve the labeling.
Our MIDI transcription files have a melody track, simplifying melody
extraction, and we quantize melodies to 16$^{th}$s. 

\subsection{Algorithm Design}
Here, we present an overview of our data and analysis algorithm for
repetition-based phrase identification. Due to space limitations, we
have posted a more detailed description at \url{cs.cmu.edu/~music/shuqid/musan.pdf}.

Given a song consisting of melody, a chord analysis, and a time signature, our goal is to determine the phrase structure with the shortest structure description length ($SDL$). The algorithm consists of:
(1) Find all matched phrase pairs (repetitions) of 
 equal-length and non-overlapping song segments of length 4 to 20 measures. 
 (2) Merge matching pairs into sets of matching phrases.
 If we view each of the phrases in our matched phrase pairs as a node
 in an undirected graph and add edges between the phrases that are
 matched, then finding all the sets is equivalent to finding all 
 maximal cliques in this sparse undirected graph. We call these \textit{phrase sets}.
 (3) Find the best structure minimizing \textit{SDL}.
The problem is equivalent to the maximum weighted clique problem 
in the undirected graph. Since this is an NP-complete problem, 
we combine dynamic programming, A* search, and heuristics to create 
a good solution with reasonable efficiency. 

Since even long songs have only hundreds of measures and the number of phrase sets grows roughly linearly with song length, computation is feasible. 
Our full algorithm correctly produced 92\% of the human-labeled structures. The average run time of each song on a laptop with a 2.3GHz 8-Core Intel Core-i9 and a 64GB-2667MHz-DDR4 RAM is 345$\thinspace$s, but for 80\% of the songs, the average run time is only 21$\thinspace$s.

\section{Hierarchical Structure Exploration}\label{sec:stucture_exploration}
In this section, we characterize the lower-level phrase structure and the higher-level section structure we found in our data set.

\subsection{Phrase-level Structure statistics}

\begin{figure}[hbt!]
\vspace{-10pt}
\setlength{\abovecaptionskip}{0pt} 
\setlength{\belowcaptionskip}{0pt}
\begin{minipage}[b]{0.5\textwidth}
 \centerline{
 \includegraphics[width=0.95\columnwidth, height=0.65\columnwidth]{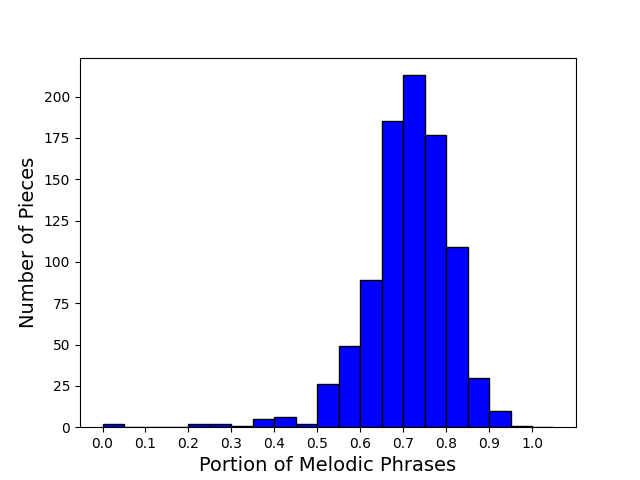}}
 \caption{Distribution of proportion of \newline
repeated melodic phrases.}
 \label{fig:main_por}
\end{minipage}
\begin{minipage}[b]{0.5\textwidth}
  \centerline{
 \includegraphics[width=0.95\columnwidth, height=0.65\columnwidth]{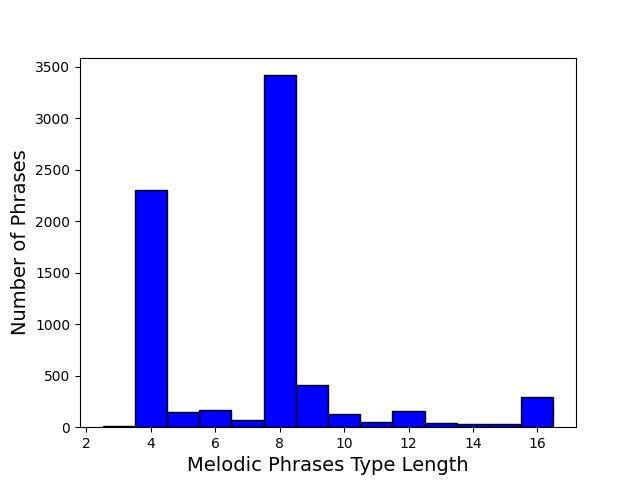}
 }
 \caption{Distribution of melodic phrase length.}
 \label{fig:main_length_distr}
\end{minipage}
\vspace{-12pt}
\end{figure}

What portion of the song is covered by repetition structure? Figure \ref{fig:main_por}
shows that in most songs, repeated melodic phrases cover 50\% to 90\% of the whole song.

Figure \ref{fig:main_length_distr} shows the distribution of different phrase lengths among phrases. The majority of melodic phrases have 4 or 8 measures (but we consider longer, higher-level sections below).

\subsection{Higher-level Sectional structure}
The importance of multi-level hierarchy in music is firmly established. Structure in traditional forms ranges from sub-divisions of beats to movements of symphonies. We looked for automatic ways to detect structures above the level of our ``phrases,'' which are based on repeated patterns. One indication of higher-level structure is the presence of non-matching
({\tt X}) and non-melodic phrases that partition the song structure.
In our analysis, successive non-melodic phrases and {\tt X} phrases with total lengths of more than two measures indicate the boundaries of high-level \textit{sections}. For example, the song with structure analysis\newline
{\tt i4A8B8x4A8X2B8B8c4c4B9o2}, after removing separator phrases {\tt i4}, {\tt x4}, {\tt c4c4}, and {\tt o2}, has three sections: {\tt A8B8}, {\tt A8X2B8B8} and {\tt B9}. For lack of more standard terminology, we refer to our low-level repetition segments as \textit{phrases} and these higher-level segments as \textit{sections}.

\begin{figure}[hbt!]
\vspace{-10pt}
\begin{minipage}[t]{0.32\textwidth}
 \centering
 \includegraphics[width=0.95\columnwidth, height=0.65\columnwidth]{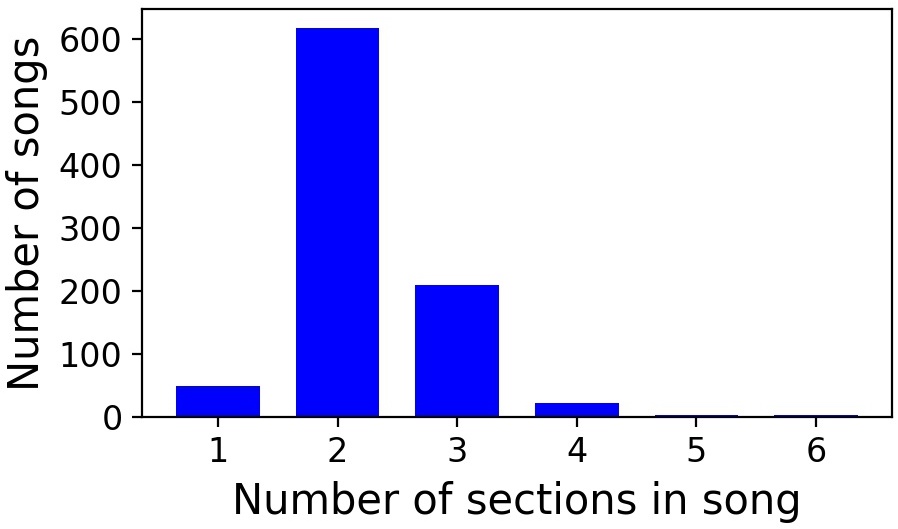}
 \caption{Distribution of\newline
 the number of high-level\newline
 sections in a song.}
 \label{fig:num_section_distribution}
\end{minipage}
\begin{minipage}[t]{0.32\textwidth}
  \centerline{
 \includegraphics[width=0.95\columnwidth, height=0.65\columnwidth]{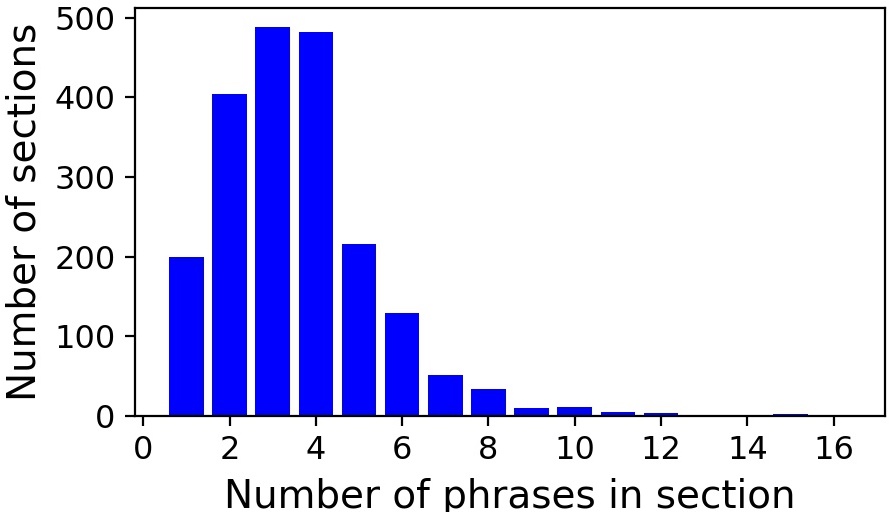}
 }
 \caption{Distribution of\newline
 the number of phrases\newline
 in a section.}
 \label{fig:phrase_num_section_distribution}
\end{minipage}
\begin{minipage}[t]{0.32\textwidth}
  \centerline{
 \includegraphics[width=0.95\columnwidth, height=0.65\columnwidth]{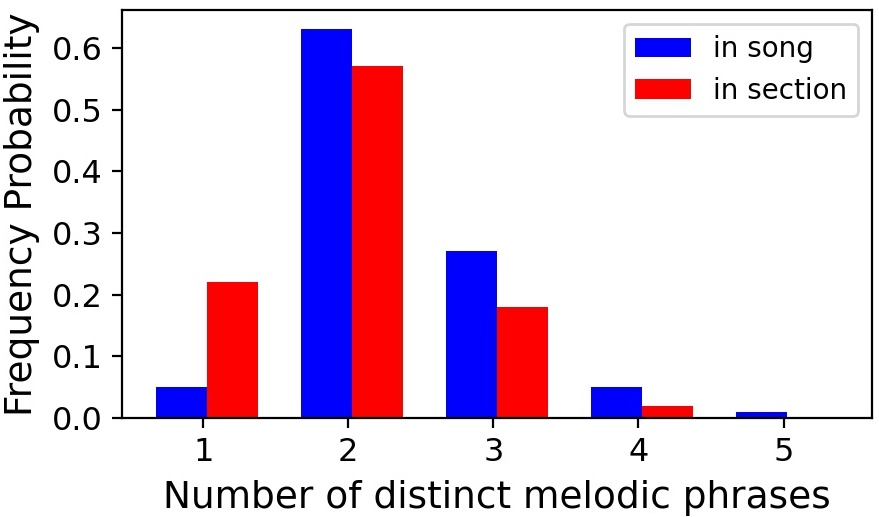}}
 \caption{Distribution of the number of distinct melodic phrases.}
 \label{fig:num_different_repetition_distribution}
\end{minipage}
\vspace{-12pt}
\end{figure}

We found that most of the songs have 
two or three sections (Figure \ref{fig:num_section_distribution}), 
and each section typically has 1 to 6 phrases (Figure \ref{fig:phrase_num_section_distribution}).  
Over 90\% of songs have 
two or three distinct \textit{phrases} with melody (e.g. {\tt A}, {\tt B}, ...). Within each section in the song, there are typically one to three distinct melodic phrases (Figure \ref{fig:num_different_repetition_distribution}).

Data further shows that 20\% of sections are exact repetitions of the previous section in terms of phrases; 29\% of the successive sections repeat a suffix of the previous section (e.g. {\tt AAB AB}) while 18\% repeat a prefix (e.g. {\tt ABB AB}).


\section{Interactions with Segment Structure}\label{sec:harmonic_analysis}
We could have used any number of ways to form 
higher-level structure (sections), but our choice is supported
by the finding of interactions between sections, melody, harmony 
and rhythm that do not occur
so strongly at the phrase level, suggesting that the section 
structure is not just an arbitrary construction. 

\newcommand{\Chd}{\text}

\begin{figure}[hbt!]
 \centerline{
 \includegraphics[width=0.8\columnwidth, height=5.6cm]{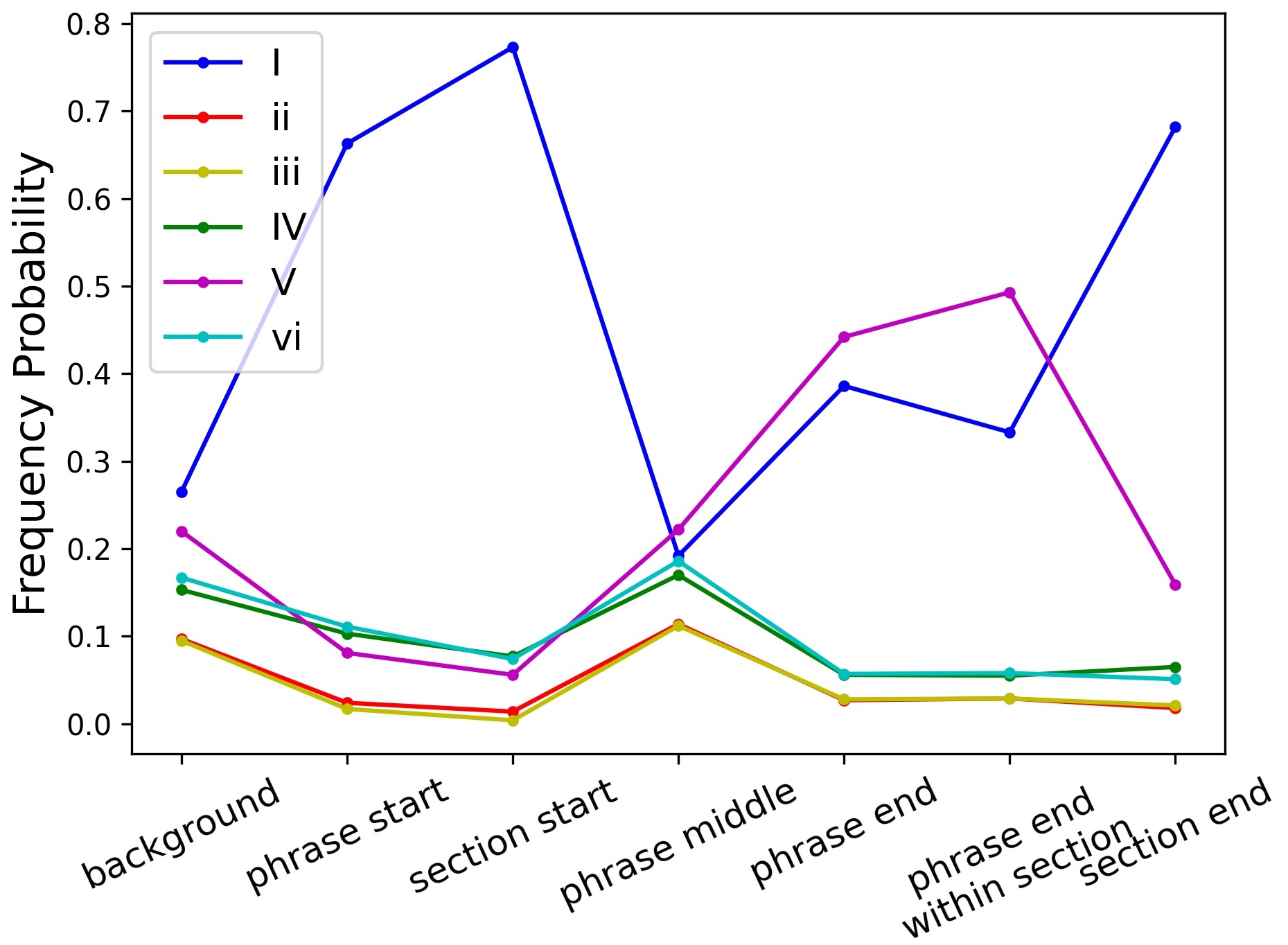}}
 \caption{Chord frequency probability at different level of structure in major mode. X-axis represent different locations in phrase and sections. \textit{Background} means no location constraint, for comparison.}
 \label{fig:chord_histogram_structure}
 \vspace{-12pt}
\end{figure}

In Figure \ref{fig:chord_histogram_structure}, we show 
probabilities of different harmonies at different locations in phrases 
and sections in major mode. We are much more likely to see a {\Chd I} 
at the beginning of a phrase and at the end of a section. {\Chd I} and {\Chd V} chords are more popular at the ends of phrases (about equally).
We expected to see a predominance of {\Chd I} chords at the ends 
of phrases, but as the last two categories reveal, the {\Chd V} 
is a more common
ending \textit{within} a section, while the {\Chd I} chord is more 
common at the \textit{end} of a section. Here, we see 
significant interactions not only between structure and harmony 
but between \textit{different structural levels}.
We evaluate the significance of these differences by assuming a null hypothesis of equal probability everywhere (the \textit{background} category in Figure \ref{fig:chord_histogram_structure}) and using one-tailed unpaired t-tests. All the test results are significant ($P < .0001$). 

Chord \textit{transitions} at the ends of phrases or sections proved 
to be significantly different from general transition probabilities 
at other positions in the phrase. For example in major mode, 58\% of
progressions 
at the \textit{end} of the section are {\Chd V--I} (Authentic Cadence in music theory). Transition probabilities from {\Chd V--I} at the end of phrase, end of phrase in the middle of a section, and end of section are 0.89, 0.84 and 0.94, while the average transition probability in all other positions is only 0.47.  

\begin{figure}[hbt!]
 \centerline{
 \includegraphics[width=0.8\columnwidth, height=5.9cm]{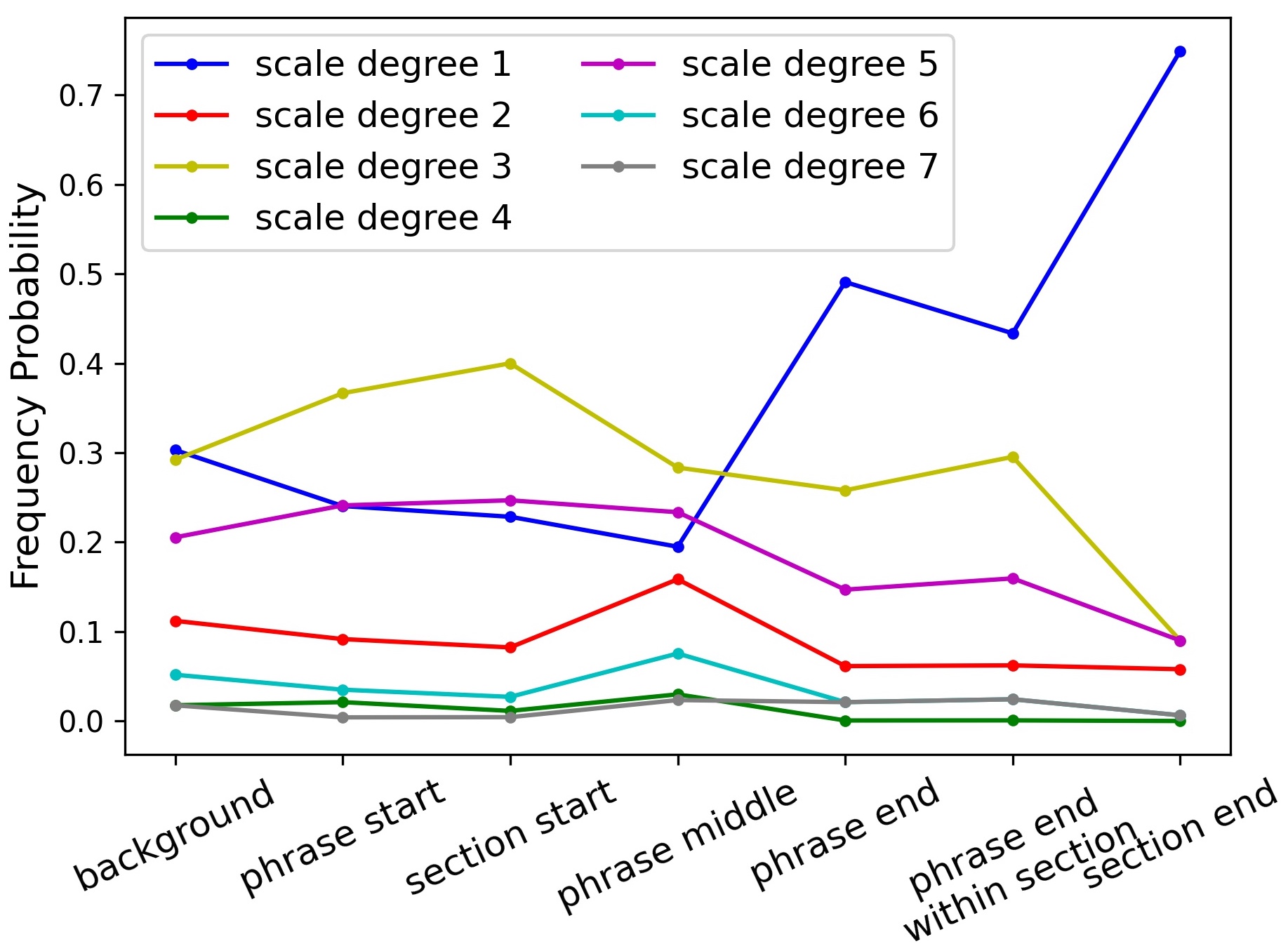}}
 \caption{Melody pitch distribution probabilities conditioned on {\Chd I} chord at different levels of structure in major mode.}
 \label{fig:pitch_Ichord_structure}
 \vspace{-12pt}
\end{figure}

Phrase and section structures also influence the distribution of 
melody pitches. 
Figure \ref{fig:pitch_Ichord_structure} shows probabilities of different melody pitches at different locations in phrases and sections, counting only pitches in the context of a {\Chd I} chord in the major mode.
Scale degree $\hat{1}$ in melody tends to occur at the \textit{end} of sections, but not at the
\textit{start} or \textit{middle} of phrases. While scale degree $\hat{3}$ is common
in the start of phrases, but not at the end of sections.

We are also interested in the duration and rhythm of the melody. 
The distribution of note length at the beginning and middle of phrases is about the same as the overall distribution, consisting mostly of short notes. In contrast, the phrase endings mostly consist of longer notes. We also observed a difference in phrase endings depending on position. For example, only 6.4\% of whole-or-longer notes occur at the ends of phrases in the \textit{middle} of a section, while 72\% occur at the ends of phrases at the \textit{end} of a section. 

We also discovered two measures of phrase harmonic structure that correlate with year of composition. (Our data spans 7 decades of music.)
Although space limits a complete discussion, we found cross-phrase similarity decreases with date (contrast between song sections increases) ($P < .01)$, and phrase complexity (in an information theoretical sense) \textit{increases} with date, indicating generally longer phrases and more variety of chord types ($P < .002$).
\section{Discussion and Future Direction}\label{sec:discussion}
The data-driven analysis results in this paper show that music 
elements such as harmony, melody and rhythm behave differently 
at different positions relative to the hierarchical music structure. 
These music-structure-related features support many aspects of traditional music theory. For example, in our analysis, half 
cadences are more often seen at the ends of phrases, but only 
in the middle of sections, consistent with the music theoretic 
concept that a half cadence calls for continuation. 
It is worth noting that the phrase structure extraction algorithm is fully based 
on repetition and similarity without using any knowledge of these music 
concepts. Thus, our approach forms a good test for music theory and existing 
domain knowledge.

The algorithms we proposed for extracting hierarchical repetition structures from MIDI files have a high accuracy of 93\% compared to human labeling, and can be used to analyze other MIDI data sets. Our findings can guide music imitation or generation and can also be used to evaluate whether songs follow structural conventions. Notice that in the phrase-level structure analysis algorithm, parameters are manually tuned, but perhaps they could be adjusted automatically according to different styles of music. 

Future work might strive to learn more about variations between similar phrases and how contrasting phrases are constructed. It would also be interesting to compare other data sets, including non-pop music. We have only begun to look for interactions between structure, melody, harmony and rhythm, and these initial results show this to be a promising research direction. The idea that structural tendencies change over time is also promising.

Our results with Chinese pop music are consistent with basic concepts of Western music theory, so we suspect that similar results would be obtained with Western pop music. Still,
it would be interesting to conduct a comparative study with Western pop songs. Future work might also investigate more robust indicators of sections. It seems that the non-melodic phrases we use to detect sections are not present in all styles. Consider a repeated form such as {\tt AABA|AABA}. There might be ways to identify these higher-level sections which are not separated by non-melodic phrases.



\section{Conclusion}\label{sec:conclusion}
We believe this is the first study to analyze connections between different levels of music structure and the elements of harmony, melody and rhythm using a data-driven approach. We introduced a new hierarchical structure analysis algorithm. With it, 
we analyzed harmony, melody and rhythm in the context of multi-level structure.
This work suggests there is 
still much to be learned about the role of structure in music, and that we can use hierarchical structure to inform future work on music style, analysis, evaluation and generation.

Our data set, annotations, and experimental results are released at: \url{https://github.com/Dsqvival/hierarchical-structure-analysis}.


\bibliographystyle{apacite}
\bibliography{paper.bib}

\end{document}